\RequirePackage{amsmath}
\documentclass[runningheads]{llncs}
\usepackage[utf8]{inputenc}
\usepackage[english]{babel}
\usepackage{amssymb}
\usepackage{bbm}
\usepackage[strings]{underscore}
\usepackage{xcolor}

\usepackage{graphicx}
\usepackage{booktabs} 
\usepackage{multirow}

\usepackage{listings}
\usepackage{hyperref}
\usepackage{color}
\usepackage{soul}

\newcommand{\argmin}{\ensuremath{\operatorname{arg\,min}}}
\def\HiLi{\leavevmode\rlap{\hbox to \hsize{\color{green!50}\leaders\hrule height .8\baselineskip depth .5ex\hfill}}}

\usepackage[ruled,vlined,linesnumbered]{algorithm2e}
\SetKwInOut{Parameter}{Parameter}

\makeatletter
\g@addto@macro\normalsize{%
\setlength\abovedisplayskip{2pt}
\setlength\belowdisplayskip{2pt}
\setlength\abovedisplayshortskip{2pt}
\setlength\belowdisplayshortskip{2pt}
}
\makeatother

\def\EQ#1{\begin{eqnarray}#1\end{eqnarray}}

\newcommand{\ket}[1]{{\left\vert{#1}\right\rangle}}

\begin{document}
\title{Tabu-driven Quantum Neighborhood Samplers}

\author{Charles Moussa\inst{1} \and
Hao Wang\inst{1} \and
Henri Calandra\inst{2} \and
Thomas B\"{a}ck\inst{1} \and
Vedran Dunjko\inst{1}}

\authorrunning{C. MOUSSA et al.}

\institute{LIACS, Leiden University, Niels Bohrweg 1, 2333 CA Leiden, Netherlands \and
TOTAL SA, Courbevoie, France
}
\maketitle              
\begin{abstract}
Combinatorial optimization is an important application targeted by quantum computing. However, near-term hardware constraints make quantum algorithms unlikely to be competitive when compared to high-performing classical heuristics on large practical problems. One option to achieve advantages with near-term devices is to use them in combination with classical heuristics. In particular, we propose using quantum methods to sample from classically intractable distributions -- which is the most probable approach to attain a true provable quantum separation in the near-term -- which are used to solve optimization problems faster. We numerically study this enhancement by an adaptation of Tabu Search using the Quantum Approximate Optimization Algorithm (QAOA) as a neighborhood sampler. 
We show that QAOA provides a flexible tool for exploration-exploitation in such hybrid settings and can provide evidence that it can help in solving problems faster by saving many tabu iterations and achieving better solutions.

\keywords{Quantum Computing  \and Combinatorial Optimization \and Tabu Search.}
\end{abstract}
\section{Introduction}
\label{intro}

\par
In the Noisy Intermediate-Scale Quantum (NISQ) era \cite{Preskill2018quantumcomputingin}, hardware is limited in many aspects (e.g., the number of qubits, decoherence, etc.), which prevent the execution of fault-tolerant implementations of quantum algorithms. Therefore, hybrid quantum-classical algorithms were designed for near-term applications. Examples include algorithms for quantum chemistry problems \cite{VQE,hardwareAnsatz}, quantum machine learning \cite{PQC} and combinatorial optimization \cite{QAOA}. They generally consist of one or many so-called parameterized quantum circuits (or variational quantum circuits), where the circuit architecture is fixed but the parameters of individual gates are adapted in a classical loop to achieve a computational objective. 
\par 
Designed for combinatorial optimization, the Quantum Approximate Optimization Algorithm (QAOA) \cite{QAOA} consists of a quantum circuit of a user-specified depth $p$, involving $2p$ real parameters. To the limit of infinite depth, it converges to the optimum. While numerous works have been studying various theoretical and empirical properties of QAOA \cite{qaoaConcentrates,Crooks2018b,qaoaperf,qaoavsannealing,qalgoselection}, many practical challenges remain. Indeed, only small-sized problems and very limited $p$ can be run on real hardware, which severely limits the quality of solution obtained empirically \cite{qaoasycamore,cvarQAOA,gibbsQAOA}. Lastly, many open questions still remain, e.g., regarding the comparison of QAOA with other heuristic methods on various cases of instances which stem from particular problem domains, with different optimizers, and with varying levels of experimental (or simulated) noise. One reason of why so many uncertainties remain is that classical simulation is computationally very expensive, and quantum devices are still scarce to prevent large real world tests \cite{simulqaoapaper,qaoasycamore}.
\par 
In contrast to optimization problems, quantum advantage has been demonstrated in sampling \cite{googlesupremacy}. Indeed, theoretical results establish quantum advantage in producing samples according to certain distributions of constant-depth quantum circuits \cite{Bravyi308}. In this direction, it has been demonstrated that the sampling of the QAOA circuit, even at $p=1$, cannot be efficiently simulated classically \cite{qaoaSupremacy}. The above considerations point to a possibility of utilizing sampling features of QAOA for neighborhood explorations with the added benefit that, since the neighbourhood may be limited to fewer variables, a smaller quantum device may already lead to improved performance of a large instance.
\par
It is interesting to delve into sampling aspects in the domain of classical local search algorithms, where we seek the optimum in the vicinity of the current solution with respect to either the original optimization problem or a subproblem thereof, using a deterministic or stochastic sampling strategy~\cite{0092410}. Such a sampling-based local procedure is typically realized by the combination of some parametric distribution family for drawing local trial points (e.g., the binomial or a power-law distribution~\cite{DoerrLMN17}) and a selection method for choosing good trial points, and hence the overall outcome of this procedure results in the family of sampling distributions~\cite{DoerrD18,DBLP:conf/gecco/WatsonJ07,Lehre2010UIOCrossover}. 
\par
In this work, we propose to use QAOA circuits as local neighborhood samplers, having a malleable support in (many) good local optima but still allowing a level of exploration (which is desirable since local optima may not lead to global optima). This introduces the topics of sampling and multiobjective aspects of QAOA that allow balancing between exploration and exploitation. To this end, we study its combination with tabu search (TS), a metaheuristic that has been successfully applied in practice for combinatorial optimization by local search. Moreover, to control the trade-off between exploration and exploitation, we add this critical component in TS to the specification of the standard QAOA circuit. 
\par 
\textbf{Contributions -} In this work, we construct an algorithm incorporating QAOA in TS with the usual attribute-based short-term memory structure (a.k.a the tabu list). With our approach, we kill two birds with one stone: we gain quantum enhancements, while the local properties of tabu search can make the required quantum computations naturally economic in terms of needed qubit numbers, which is vital in the near-term quantum era. We analyse and benchmark this incorporation with \emph{small QAOA depths} against a classical TS procedure on Quadratic Unconstrained Binary Optimization (QUBO) problems of up to 500 variables. We also propose a penalized version of QAOA incorporating knowledge from a current solution. We find that QAOA is often beneficial in terms of saved iterations, and can exceptionally find shorter paths towards better solutions. The structure of the paper is as follows. Section \ref{background} provides the necessary background on QUBOs, TS and QAOA. In section \ref{tabuSTM}, we detail the TS procedure incorporating a short-term memory structure with QAOA. The results of our simulations are presented in section \ref{simus}. We conclude our paper with a discussion in section \ref{discussion}.

\section{Background}
\label{background}

The QUBO formulation can express an exceptional variety of combinatorial optimization (CO) problems such as Quadratic Assignment, Constraint Satisfaction Problems, Graph Coloring, Maximum Cut \cite{Kochenberger2006}. It is specified by the optimization problem $ \min_{x \in \{0,1\}^n} \sum_{i \le j} x_i Q_{ij} x_j$ where $n$ is the dimensionality of the problem and $Q\in\mathbb{R}^{n\times n}$.
This formulation is connected to the task of finding so called ``ground states", i.e., configurations of binary labels $\{1,-1\}$ minimising the energy of spin Hamiltonians, commonly tackled in statistical physics and quantum computing, i.e.,:
\EQ{
 \textstyle\min_{s \in \{-1,1\}^n} \sum_i h_i s_i + \textstyle\sum_{j>i} J_{ij} s_i s_j, \label{ising}
 }
where $h_i$ are the biases and $J_{ij}$ the interactions between spins.
\par 
The QAOA approach has been designed to tackle CO problems and was inspired from adiabatic quantum computing \cite{QAOA}. Firstly, the classical cost function is encoded in a quantum Hamiltonian defined on $N$ qubits by replacing each variable $s_i$ in Eq.~\eqref{ising} by the single-qubit operator $\sigma_i^z$:
\EQ{ H_C =  \textstyle\sum_i h_i \sigma_i^z + \sum_{j>i} J_{ij}  \sigma_i^z \sigma_j^z .}
Here, $H_C$ corresponds to the target Hamiltonian and the bitstring corresponding to the ground state of $H_C$ also minimizes the cost function. Secondly, a so-called \emph{mixer Hamiltonian} $  H_B = \sum_{j=1}^N  \sigma_j^x$ is leveraged during the procedure. These hamiltonians are then used to build a layer of a quantum circuit with real parameters. This circuit is initialized in the $\ket{+}^{\otimes N}$ state, corresponding to all bitstrings in superposition with equal probability of being measured. Then, applying the layer $p$ times sequentially yields the following quantum state:
$$  \ket{\mathbf{\gamma},\mathbf{\beta}} = e^{-i\beta_p H_B} e^{-i\gamma_p H_C} \cdots e^{-i\beta_1 H_B} e^{-i\gamma_1 H_C} \ket{+}^{\otimes N},  $$
defined by $2p$ real parameters $\gamma_i,\beta_i, i=1...p$ or \emph{QAOA angles} as they correspond to angles of parameterized quantum gates. Such output corresponds to a probability distribution over all possible bitstrings. The classical optimization challenge of QAOA is to identify the sequence of parameters $\mathbf{\gamma}$ and $\mathbf{\beta}$ so as to minimize the expected value of the cost function from the measurement outcome. In the limit of infinite depth, the distribution will converge to the optima.
\par
Tabu Search (TS) \cite{Glover2013} is a meta-heuristic that guides a local heuristic search procedure to explore the search space beyond local optimality. One of the main components of TS is its use of adaptive memory, which creates a more flexible search behavior. Such framework allows using a quantum algorithm as a local search search tool, for solving large instances with limited-sized quantum devices. Various works leveraged TS for solving QUBOs \cite{Kochenberger2014,amts,MTS,GloverDiversificationdrivenTS,ITS} using short-term and long-term strategies used during the search. We note also different hybrid settings that combine a basic TS procedure with another framework such as genetic search~\cite{L2010AHM} and Path Relinking~\cite{Wang2012PathRF}. TS was also incorporated with quantum computers to tackle larger problems beyond their limitations. Indeed, finding methods to leverage smaller devices is of main importance. Many divide and conquer approaches have been designed for quantum circuits and algorithms~\cite{vedran1,vedran2,tradingclassquantumresources,simulatinglargesmall}. In this paper, the size of the QC comes into play more naturally as a hyperparameter defining the \guillemotleft radius\guillemotright\xspace of the search space.
\par 
With respect to the interplay between TS and quantum techniques, to our knowlege TS has only been considered from the perspective of D-Wave quantum annealers. The first approach of this kind is an algorithm called qbsolv~\cite{Booth2017PartitioningOP}. It starts with an initial TS run on the whole QUBO. Then the problem is partitioned into several subproblems solved independently with the annealer. Subproblems are created randomly, by selecting variables. Non-selected ones have their values fixed (clamping values) from the TS solution. The subsolutions are then merged and a new TS is run as an improvement method. The second approach is an iterative solver designed in \cite{Rosenberg2016BuildingAI}. At each iteration, a subproblem is submitted to the annealer. The subproblem is obtained by clamping values from a current solution. A tabu list is used in which each element is a list of variables of length $k$. Each element is kept tabu for a user-defined number of iterations. In contrast in this work, we consider using QAOA in combination with TS.

\section{Tabu-driven QAOA sampling}
\label{tabuSTM}

Inspired by the above-mentioned works, we use a simple TS procedure where QAOA is added in the neighborhood generation phase to solve QUBO problems. Note that we could also apply QAOA in more sophisticated frameworks, but a simpler approach is easier for understanding the benefits of QAOA with TS.
\par 
Local search algorithms explore a search space by generating sequences of possible solutions which are refined. At each step we generate so-called neighborhood from a current solution. In particular, if we denote the current solution $x$, a generated neighborhood corresponds to candidates $x^{\prime}$ that differ by at most $k$ bits. We denote this set as $N_k (x) = \{ x^{\prime} \in \{0,1\}^n | \delta_H (x^{\prime}, x) \le k \} $, where $\delta_H$ denotes the Hamming distance. For a simple one bit-flip generation strategy, this corresponds to $k=1$ and TS uses a modified neighborhood due to tabu conditions. Although increasing $k$ could help exploration, the neighborhood generation comes at exponential cost. But this could mean finding better solutions in fewer TS iterations, and thus also in principle overall faster if a fast good method for neighborhood exploration is devised.
\par 
This motivates the use of a quantum algorithm as a proxy for exploring $N_k (x)$. Specifically, we will use QAOA which $2p$ real parameters are tweaked in a continuous optimization scheme resulting in a probability distribution on $N_k$. Increasing $p$ (assuming the optima are found over the parameters) will improve the quality of the output (likelihood of returning an actual global optimum).
\par 
However, in the case of local search, a greedy strategy that tends to select the best point in the neighborhood would not only lead to potential stagnation, but also result in longer optimization time (unless the neighborhoods are already the size of the overall problem)~\cite{DBLP:books/daglib/0006199}. Indeed, one may also consider modifications which impose (various) notions of locality, which are usually not considered in standard QAOA uses where it is used for the entire instance, with the sole goal of finding optima. To this end, we first outline the basic TS procedure generally used to solve QUBO problems \cite{GloverDiversificationdrivenTS,L2010AHM,Wang2012PathRF}. Then, we show how QAOA can be combined with the latter. Finally, we propose a modification of QAOA that balances between going for the global optimum, and prioritizing local improvements relative to the current TS solution. 

\subsection{The basic TS algorithm}

The basic TS procedure for solving a QUBO with objective function $f(x)$ is described in Alg.~\ref{alg:tsoneflipQAOA}, for $k=1$ excluding the green-highlighted part. It uses a simple \emph{tabu list} recording the number of iterations a variable remains tabu during the search. A variable can be set tabu for a fixed number of iterations (denoted Tabu tenure TT) but also with a random tenure. Each iteration can be considered as updating a current solution denoted $x$, exploring a modified neighborhood $N^{\prime}_1$ due to the tabu considerations. Generally, $x$ is chosen greedily when evaluating the objective function over candidates $x \in N^{\prime}_1$. 
\par
For large problem instances, there exists an efficient evaluation technique for QUBO solvers leveraging one-bit flip move~\cite{fastincremental}. Let $\Delta_x = f(x^{\prime}) - f(x)$ be a move value, that is the effect in objective of going to $x^{\prime}$ from $x$. For one-bit flip moves, we denote as $\Delta_x (i)$ the move value upon flipping the $i$-th variable, which can be computed using only the QUBO coefficients. The procedure records a data structure storing those move values, which is updated after each TS iteration.
\par
Initially, all variables can be flipped (line 5). At each iteration, the tabu solution $x$ is updated by flipping the variable that minimizes the objective over the neighborhood obtained by one-bit flip moves over non-tabu variables (lines 6-8). If the new tabu solution improves over the best recorded solution, the aspiration is activated. In this case, the tabu attribute of the flipped variable is removed. The tabu list is finally updated (lines 18-23) and iterations continue until the stopping criterion is reached. This can be either a maximum number of TS iterations, and/or a maximum number of TS iterations allowed without improvement of the best solution (\emph{improvement cutoff}).

\subsection{QAOA neighborhood sampling}\label{STMqaoa}
In the usual TS algorithms, the neighborhood consists of candidates with Hamming distance one relative to the current tabu solution $x$. We note that sometimes considering also neighbors that are at most $k$-Hamming distance away from $x$ helps in finding better solutions. The number $k$ can be set in our case as large as the (limited) number of available qubits in a quantum hardware.
\par 
To study the exploration of such neighborhoods, a brute-force generation approach is initially tested, thereafter replaced by QAOA. As stated before, getting the optimum for subproblems in TS may lead to getting stuck during the search. QAOA, by definition, is a flexible framework as an exploration-exploitation tool. On the one hand, QAOA generates better solutions the deeper the circuit ($p$) , and the better the classical optimization procedure within QAOA is. It is known it can have advantages over various standard algorithms, e.g. Simulated and Quantum Annealing \cite{qaoavsannealing}. To extract all advantages from the capacities of QAOA, we can further modulate the distribution over outputs it produces by limited depth or, as we present next, modifications to the QAOA objective to prioritize a more local behaviour. Such flexibility is important, not only for the exploration-exploitation trade-off as it provides interesting ways fine-tune the algorithm depending on the instance to solve.
\par 
First, the choice of variables to run QAOA on needs to be adressed. Considering the $\binom{N}{k}$ possibilities would be intractable. Variables can be chosen randomly but an approach incorporating one-bit flip move values can help in guiding towards an optimum. The $k$ variables can be chosen amongst the non-tabu ones at each step. Plus, this means QAOA is an attempt at improving over the solution one would get with the one-bit flip strategy outlined previously. 
One can either select the $k$ variables greedily or add randomness by using the one-flip gains as weights for defining a probability to be chosen. For simplification, we consider the greedy selection based on one-bit flip move values. If we consider the chosen variables that were flipped, the update step of the incremental evaluation strategy can be applied. Let $l \le k$ be the number of different bits. The newly generated candidate can be considered as a result of $l$ sequential one-bit flips. Thus, $l$ calls to the above-mentioned efficient procedure are required. 
\par 
A second consideration concerns the tabu strategy for updating the tabu list. We choose to set as tabu the variables amongst the $k$ chosen ones that were flipped. Choosing to flip all chosen ones could be problematic as it could lead to all variables being tabu very early during TS. An aspiration criterion can be used if the new candidate gives the best evaluation found during the search.
\par 
Finally, the question of how to run QAOA is of main importance. In our first scenario, QAOA will be run as a proxy for brute-force (with exploration properties) to optimize the subproblem defined over the $k$ chosen variables. This is done by fixing in the QUBO the non chosen ones from the current tabu solution $x$. The depth $p$ of QAOA can be user-defined. In this work, we limit $p$ to 2 to showcase sampling aspects of QAOA at small depth.
\par 
Our QAOA-featured TS is outlined in Alg.~\ref{alg:tsoneflipQAOA} where QAOA addition is indicated in green shades. It starts with the same steps as with the standard Tabu search algorithm until line 8. The QAOA part kicks in from line 9 by firstly choosing a subset of $k$ variables, and executing QAOA on the sub-QUBO problem where we optimize over the chosen $k$ variables while keeping the remaining bits the same with the current point $x$. After obtaining the best point from QAOA (lines 11-13), we select the better one from the QAOA outcome $x^{k\text{-bit}}$ and the best one-bit flip point $x^{1\text{-bit}}$ and use it to update the current search point. The move values are then updated by $l$ calls of the fast incremental method (line 14). Finally, if the best-so-far point is improved by the updated search point, we drop the accepted bit flips from the tabu list (line 21). Otherwise, we reset their tabu value to the sum of tabu tenure and a random tenure (line 23).

\begin{algorithm}[!t]
\label{alg:tsoneflipQAOA}
\begin{small}
\KwIn{An initial solution $x_0$, Cost function $f(x)$}
\Parameter{Tabu tenure TT, Random tabu tenure rTT, subproblem size $k$}
\KwOut{The best solution achieved $x^{\ast}$}
 $x^* \gets x \gets x_0$\; 
 $Tabu(i) \gets 0, \Delta_x(i) \gets 0$ for $i \in [1..N]$\;
 \While{stopping criterion not reached}{
  $x^{\text{pre}} \gets x$\;
  \For{$ i \in [1..N]\cap Tabu(i) = 0$}{
    $x^{(i)}_i \gets  1 - x_i, x^{(i)} \gets x$\;
    one bit-flip gain: $\Delta_x(i) = f(x^{(i)}) - f(x)$\;
  }
  $x^{1\text{-bit}} \gets  x^{(j)}, \; j \gets \argmin\Delta_x(i)$\;
  \HiLi Select greedily or randomly a subset of variables $K \subseteq I$ s.t. $|K| = k$\;
  \HiLi Get a new QUBO by fixing the $N-k$ other variables in $x$\;
  \HiLi Run QAOA and get the best sample $\hat{x}$ minimizing the new QUBO\;
  \HiLi $x^{k\text{-bit}}_i \gets x_i$ for $i \in I \backslash K$ and $x^{k\text{-bit}}_i \gets \hat{x}_i$ for $i \in K $\;

  \HiLi$x\gets$ the better out of $x^{\text{1-bit}}$ and $x^{k\text{-bit}}$\;
  Update move values $\Delta_x(i)$ for $i \in [1..N] \cap x_i^{\text{pre}} \neq x_i $\;
  $aspiration \gets \text{False}$\;
  \If{$f(x) < f(x^{\ast})$}{
   $x^{\ast} \gets x$, $aspiration \gets \text{True}$\;
   }
   $Tabu(i) \gets Tabu(i) - 1$ for $i \in [1..N]\cap Tabu(i) > 0$\;
   \For{$j \in [1..N] \cap x_j^{\text{pre}} \neq x_j$}{
   \eIf{aspiration}{
    $Tabu(j) \gets 0$\;
   }{
    $Tabu(j) \gets \operatorname{TT} + \operatorname{Random}(\operatorname{rTT})$\;
   }
   }
 }
 \caption{QAOA-featured Tabu Search for solving QUBO.}
 \end{small}
\end{algorithm}

\subsection{Enforcing locality with penalized QAOA}\label{penaltyqaoa}
As a tool in local search algorithms, QAOA may be useful with modifications which impose notions of locality. We incorporate these notions in the cost hamiltonian so that they are captured during the QAOA evolution. This can be done through the cost function by adding a penalty term. A possibility is to consider the Hamming distance with a current tabu solution $x$.
Hence, the objective for QAOA becomes:
\EQ{
 \textstyle\min_{x'} [f(x') + A \delta_H(x',x)], \label{penalqubo} 
 }
where $\delta_H$ corresponds to the Hamming distance and $A$ is a constant. The right-hand side additive term of Eq.~\eqref{penalqubo} aims to encourage the output of candidates that differ by few bits from the current solution if $A>0$, and vice-versa. As a neighborhood sampler, the penalty may help in enforcing locality. However, setting the parameter $A$ is non trivial for activating the effect of the extra term.
\par 
There is also a possibility to add information about the fitness gain in differing from $x$. If switching a bit is an improving move, it would be prioritized. Our algorithm uses the one-bit flip move values $\Delta_x$ in order to select the variables to run QAOA on, from which we can construct a weighted penalty term:
\EQ{ \textstyle-\frac{1}{2} \sum_{j=1}^N \Delta_x(j) (-1)^{x_j} x_j'. \label{hammingFlipGain} } 
For a minimization problem, $\Delta_x(j) < 0$ characterizes encouraging flipping the $j$-th variable in new candidates. For a candidate in this case, $\Delta_x(j)$ is added to the cost. Conversely, $\Delta_x(j) > 0$ would result in penalizing candidates with the $j$-th bit value flipped. One can also multiply by a positive constant A for enforcing more the locality effects. 
The penalty translates in an additional operator term $H_{\text{penalty}}$, following an application of a usual cost operator in QAOA, where: 
\EQ{ \textstyle H_{\text{penalty}} = -\frac{1}{2} \sum_{j=1}^N \Delta_x(j) (-1)^{x_j} \sigma_z^j \label{hammingHamiltonianFlipGain} } 
The corresponding quantum circuit of depth one is very simple and given by $\bigotimes_i R_Z ( (-1)^{x_i} \gamma \Delta_x(i)),\gamma \in \mathbb{R}.$  

\section{Simulations}
\label{simus}
We performed extensive simulations over instances of QUBO problems publicly available in the well-know OR-Library~\cite{amts,orlib,bqpgka}. In designing them, our objectives are 1) investigating whether exploring larger neighborhood can facilitate faster convergence (in terms of TS iterations), 2) elucidating the effect of locality on QAOA output given by the introduced penalty in Eq.~\eqref{hammingHamiltonianFlipGain}, and 3) studying the utility of QAOA as a proxy for brute-force. 

\subsection{Larger neighborhood exploration benefits}

To study the first objective we replaced QAOA by brute-force search in Alg.~\ref{alg:tsoneflipQAOA}. Starting from the all-zero initial solution, we first run the basic TS with different constant tabu tenures (rTT = 0). Then, we do the same with brute-force TS for different values of $k$ up to 20. We study two regimes that differ in how TS search results, namely when $k$ is not comparatively small to $N$ on instances where $N=20$, and when it is on instances of 100, 200, and 500 variables. 
\vspace{-0.5cm}
\subsubsection{$k/N$ relatively large}

In the first regime, we assume that we can explore a large percentage (more than $25\%$) of the instance size greedily. As instances, we take the first eight instances of \emph{bqpgka} (named 1a-8a consisting of 30-100 variables), for which we solve by TS. Then we select randomly 20 variables out of $N$ and clamp values of the non-selected ones from the solutions. We do so five times per instance, resulting in 40 instances of size $20$. Then TS with different $k/N$ values for subproblems ($0.9, 0.75, 0.5, 0.25$) were tried on this suite. When using brute-force, we set $TT=2$ and tried many values for the basic TS. 
\par
\begin{figure}[!ht]
\centering
\includegraphics[width=0.81\textwidth, trim=0mm 3mm 0mm 0mm, clip]{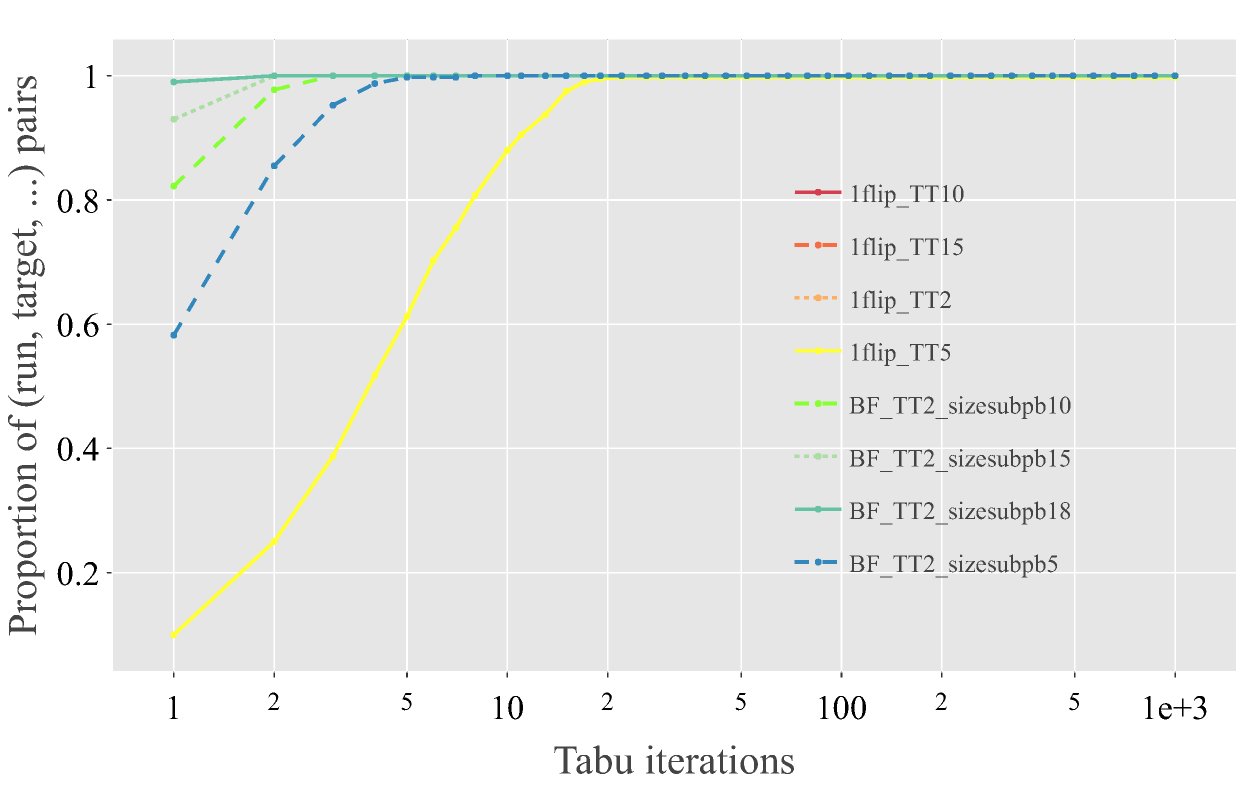}
\caption{
\small Empirical cumulative distribution function (ECDFs) of tabu iterations for each algorithm aggregated over all 20-variable problems with 10 target values evenly spanning the range of all observed function values.}
\label{size20}
\vspace{-0.8cm}
\end{figure}

 
Fig.~\ref{size20} shows the proportion of (run, target value) pairs aggregated over all functions for 10 targets generated by linear spacing using the benchmarking and profiling tool IOHprofiler~\cite{IOHprofiler} for iterative optimization heuristics. The target values were normalized by the optimum of the problems. We observed that for $k/N \ge 0.5$, these instances are straightforward to solve (in 3 iterations). The case $k=5$ required 8 iterations on one instance but managed to achieve optimality on all of them. However, the basic TS, run for 20000 iterations, failed to solve the same instance. Letting this instance aside, 5 iterations would be required for $k=5$, and 22 for the classical TS procedure. Hence, we clearly observe, as expected, degrading performances as $k/N$ decreases. This also enabled us to confirm numerically that a flip-gain based approach when considering subproblems is in general beneficial towards solving QUBOs.

\subsubsection{$k/N$ relatively small}
In this regime, we study the case when $k/N$ is relatively small (less than $20\%$). The simulations are carried out on 15 QUBO instances from bqpgka: 100-variable (named 1d-5d), 200-variable (1e-5e) and 500-variable problems (1f-5f). Algorithms are run until the best objective value found in \cite{amts} is reached or a maximum number of TS iterations is reached.
\par 
Table \ref{res-tabledef} in Appendix display results for $k=10,15,20$ and for $TT=5,10,15$. For the basic TS, we limit the number of iterations to 20000, 1000 for the brute-force approach (200 for $k=20$ though). In these cases, we observe that we can find a $k$ such that better solutions are found using less iterations. In general, $TT$ should not be set large when exploring larger neighborhood. For instance, the case $TT=15$ started to favor the basic TS in terms of target achieved. 
\par
We also run the algorithms with different $TT$ values ranging from 2 to 10 and adding 15. Table~\ref{res-tablemanyTT} shows for the different values of $TT$, which ones achieved the best performances in terms of target only. In these cases, we observe that we can find a $k$ such that better solutions are found using less iterations, especially on the most dense instances 4f and 5f. These instances, when considering the underlying graph given by the coefficients connecting different variables, have a density of respectively $0.75$ and $1$ (a non-zero coefficient for each pair of variables). For $k=20$ the proposed approach achieved optimality where the basic TS failed. Again, we observe performances, in terms of target achieved, depend on setting well the tabu tenure in accordance with $k$. Intuitively, one could think that the larger $k$, the smaller $TT$ has to be to save iterations and achieve a better objective value. But we clearly observe counter-examples.

\begin{table}[!ht]
\caption{Best values of tabu tenure (ranging from 2 to 10, adding 15) achieving the optimum obtained with the first TS iteration(s) to reach the corresponding maximum given in \cite{amts}. The best performances per instance are highlighted in bold. The mention All means all $TT$ values reached the same solution. The NA mention means no run returns the optimum, with the best value obtained in parenthesis.}
\label{res-tablemanyTT}
\begin{center}
\begin{scriptsize}
\begin{sc}
\vspace{-0.5cm}
\begin{tabular}{l|l|l|l|l|l}
\toprule

Algo. & 1d (6333) & 2d (6579) & 3d (9261) & 4d (10727) & 5d (11626) \\
\midrule
Basic & 15 / 300 & All / 71 & <9 / 90 & 2 / 67 & 6 / 171\\
\cmidrule{2-6}
$k=10$  &  All / 13 & >2 / 61 &  6 / 39 & 2 / 31 & All / 19 \\
\cmidrule{2-6}
$k=15$  &  All / 10 & \textbf{2+5 / 26} &  All / 8 & 3 / 18 & 5 / 42 \\
\cmidrule{2-6}
$k=20$  &  \textbf{All / 7} & 5 / 37 &  \textbf{All / 5} & \textbf{All / 11} & \textbf{2 / 18} \\
\bottomrule

Algo. & 1e (16464) & 2e (23395) & 3e (25243) & 4e (35594) & 5e (35154) \\
\midrule
Basic & 9 / 419 & 6 / 493 & 10 / 190 & 15 / 170 & 6 / 238 \\
\cmidrule{2-6}
$k=10$ &  8 / 485 & 7 / 111 & All / 21 & 6 / 55 & All / 25 \\
\cmidrule{2-6}
$k=15$ &  \textbf{All / 13} & \textbf{6+7 / 31} &  \textbf{All / 13} & 3 / 41 & \textbf{2 / 20} \\

\cmidrule{2-6}
$k=20$ &  2 / 35 & NA (23370)/ 82 & 3 / 44 & \textbf{3 / 25} & 4 / 31 \\
\bottomrule

Algo. & 1f (61194) & 2f (100161) & 3f (138035) & 4f (172771) & 5f (190507) \\
\midrule
Basic & 8 / 970 & 7 / 795 &  4 / 449 & NA (172734) / 1148 & NA (190502) / 647\\
\cmidrule{2-6}
$k=10$  &  \textbf{15 / 213} & 7 / 337 & 15 / 77 & NA (172449) / 110 & NA (190502) / 126 \\
\cmidrule{2-6}
$k=15$  &  8 / 225 & \textbf{8 / 173} &  15 / 139 & NA (172734) / 46 & NA (190502) / 383 \\
\cmidrule{2-6}
$k=20$  & NA (61087) / 184 & NA (100158) / 101 &  \textbf{4 / 57} & \textbf{4 / 57} & \textbf{9 / 150}  \\

\bottomrule
\end{tabular}
\vspace{-0.1cm}
\end{sc}
\end{scriptsize}
\end{center}
\end{table}

\par
\begin{figure}[!ht]
\centering
\includegraphics[width=0.84\textwidth, trim=0mm 3mm 0mm 0mm, clip]{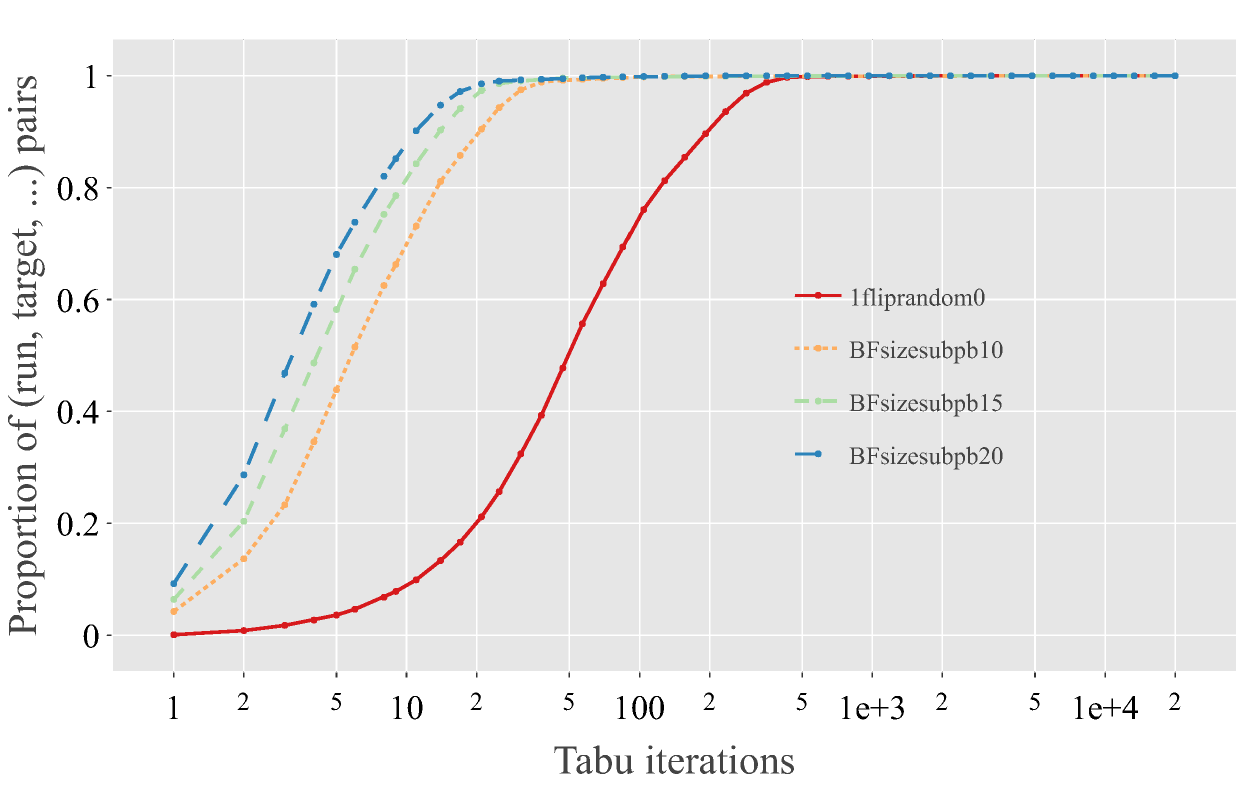}
\caption{
\small Empirical cumulative distribution function (ECDFs) of tabu iterations for each algorithm aggregated over all dimensions, all problems, and 1000 target values evenly spanning the range of all observed function values.}
\label{manyTTaggregated}
\end{figure}

Larger $k$ exploration, in this regime, turns out to not always be beneficial. This seems conter-intuitive when considering a target objective only, on an instance to instance basis comparison. Fig. \ref{manyTTaggregated} shows the proportion of (run, target value) pairs aggregated over all functions for 1000 targets generated by linear spacing. The target values were normalized by the optimum of the problems. Again, we observe that, with larger $k$, the proportion of successes is higher, when measured at the same number of iterations. Note that this can only be observed for $k=20$ up to 200 iterations. The proportion for the basic TS was close to $0.9$ while the brute-force approach was superior to 0.99. Hence, we can reach very good solutions with less iterations as $k$ increases.
\par
In summary, as opposed to the previous regime, the structure of the problems becomes very important that we have to look at performances in an aggregated way to witness the benefits of exploring larger neighborhood.  Having outlined some performances given by the brute-force approach on subproblems, we switch to QAOA, and study its sampling effect as a proxy.

\subsection{QAOA as a proxy for brute-force}
The second part of our simulations studies the output of QAOA as a proxy for brute-force. To this end, we first study an example TS run from our previous simulations. We take the subproblem QUBOs obtained at each step (except the first one), and run QAOA at $p=1$ and $2$, and we study the distribution of the energy given by $\ket{\gamma, \beta}$, after optimization, with and without the penalty term. 
\par 
Having outlined the properties of the QAOA output, we run  Alg.~\ref{alg:tsoneflipQAOA}, and study its performances in comparison to the basic TS. From the optimized angles, we try different sampling strategies to generate a candidate per iteration: just sampling once, sampling 10 times and choosing a candidate greedily, and finally consider all samples (even during optimization) greedily. The latter corresponds to a quasi brute-force (BF) approach. 

\vspace{-5mm}
\subsubsection{Energy distribution of QAOA} 
\begin{figure}[!ht]
\begin{center}
\includegraphics[width=\textwidth, trim=0 13mm 0mm 0, clip]{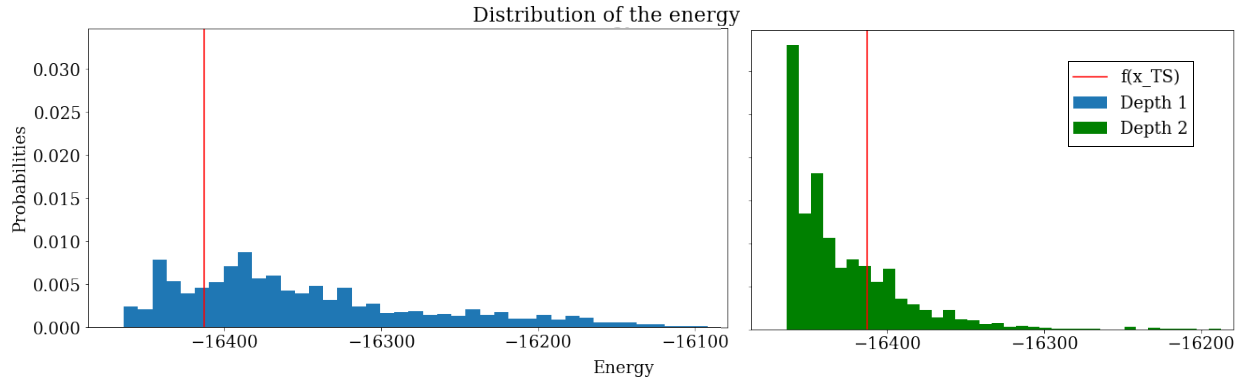}
\includegraphics[width=0.65\textwidth, clip]{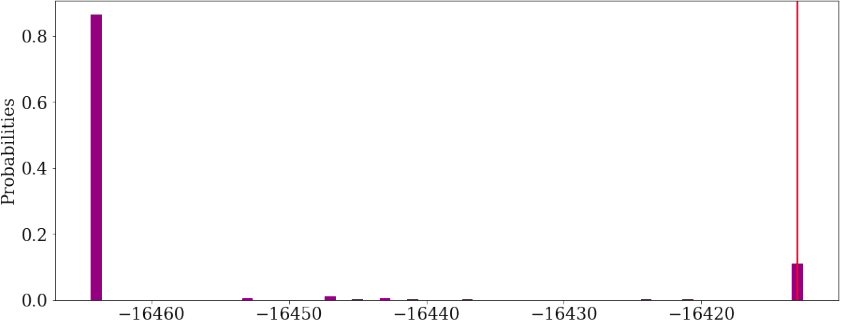}
\caption{QUBO evaluation distribution given by sampling $10^{5}$ times from the QAOA quantum state, and by running simulated annealing 1000 times (bottom), for the last TS iteration done on instance 1e.}
\vspace{-0.3cm}
\label{energy_dist}
\end{center}
\end{figure} 
As a first step, we study how the QAOA output distribution looks like at small depth, with the purpose of elucidating how it can help avoiding detrimental greedy search behavior. We consider, as an example, instance 1e for which $k=15, TT=5$ used 13 iterations greedily. The subproblem QUBOs are kept and we run QAOA on them as previously stated. The third and last iteration are interesting when considering the penalty term. The former is a case where the optimum is obtained by flipping all bits except one and where all flip moves are favorable. The last iteration has flip moves from the current tabu solution discourage flipping all bits. Plus, very few candidates ($0.2\%$) improve over the tabu solution. For these iterations, we look at the quantum state given by QAOA and analyse the distribution of the QUBO evaluation (or energy in an Ising context).
\par 
Fig. \ref{energy_dist} shows the distribution given by $10^5$ samples from the last iteration's quantum state at $p=1$ and $2$. At $p=1$, we observe a homogeneous spread with two major humps on the left and right side of the tabu solution evaluation. There is a probability of $23.9\%$ of improving from it by the quantum state. At $p=2$, we see the distribution being shifted to lower energies, yielding an improvement probability of $75\%$. The average energy is 16350 for QAOA $p=1$ and 16428 for $p=2$. Moreover, the standard deviation of the output decreases from 79.2 to 42. This is expected as to the limit of infinite depth, QAOA converges to the optimum with less variance.
\par 
When running simulated annealing 1000 times with a temperature of 17.5 using 100 steps, we observed that unlike QAOA, the energy spread is restrained to a few points, the optimum being most present. Decreasing slightly the temperature or the number of steps would always yield the optimum. However, in terms of exploration opportunities, QAOA could allow visiting different paths that may lead to fewer iterations required towards improved solutions. 
\par 
Theory shows that increasing the depth would permit QAOA to find the optimum assuming optimal parameters are found. But by limiting the depth, we can control how other good candidates are spread from the optimum. This could engender new paths to solve a problem differently, where a suboptimal solution on a subproblem leads to an easier one for QAOA towards better candidates. Note this can be done in different ways. One way would be using brute-force and perturbating or mixing the solution, which is not efficient. We could also have the same effect with simulated annealing. But in many cases, we can fail finding the optimum, and even less finding a bunch of candidates around it. Finally, due to its flexibility, QAOA permits to leverage modifications to introduce locality notions in a multiobjective scheme for local search, such as the previously mentioned penalty. In the following, we study its effect on the QAOA distribution.

\vspace{-5mm}
\subsubsection{Penalty effect}

In section \ref{penaltyqaoa}, we introduced a penalty term to impose notions of locality in QAOA as a local search tool. An extra operator based on the hamiltonian given in Eq.~\eqref{hammingHamiltonianFlipGain}, translates to a circuit of depth one concatenated with a QAOA layer. We study its effect on the QAOA distribution obtained on the resulting sub-qubos at the third and last iteration. 
\par
On iteration $3$, for both QAOA and its penalized version, the distribution, shown in Appendix Fig.~\ref{qaoa_penalized_dist_greedy}, tends to output candidates with largest Hamming distances to the current point as expected. Also, the most likely candidate is the one that completely differs from the current point, which is more favored by the penalty effect. No significant changes were observed when penalizing at $p=2$. 
\par
\begin{figure}[!ht]
\begin{center}
\includegraphics[width=\linewidth, trim =0 0mm 0 0,clip]{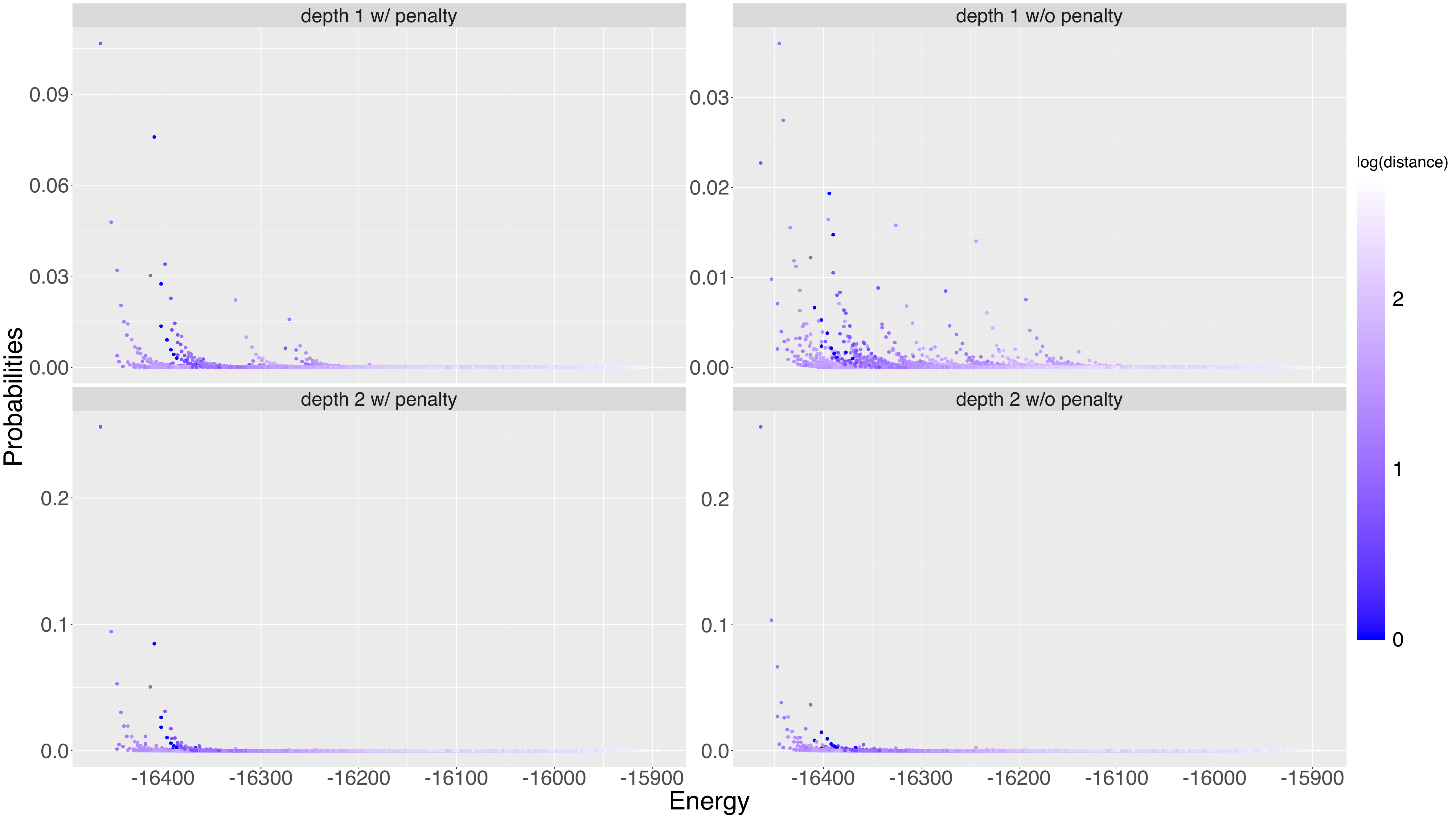}

\caption{ \small
Distribution of the evaluations (or energies) for the last iteration obtained on instance 1e given by the QAOA output state, with and without the penalty term. A colormap is given for the Hamming distance with the current tabu solution.}
\label{qaoa_penalized_dist}
\end{center}
\end{figure}

 
For the last iteration, Fig.~\ref{qaoa_penalized_dist} shows that the original QAOA at $p=1$ results in many probability peaks compared to the penalized version, which evolve to a major peak with an increased depth. The penalized version demonstrates two major peaks, from which the optimum and a close candidate to $x$ are preferred.
\par 
This characterizes the interplay between optimizing and penalizing. At $p=1$, the penalized version has a better probability of improving $x$ ($0.34$ vs $0.239$) and a higher probability of finding the optimum ($0.1$ vs $0.02$). However, the unpenalized $p=2$ version was more likely to output the optimum (respectively $0.75$ and $0.26$, where the penalty at $p=2$ yields $0.62$ and $0.26$).
\par 
In summary, using the penalty creates a balance between the greedy approach and one-bit flip gains knowledge from the current solution. This could result in smoothening the distribution while favouring interesting candidates for both objectives. This will modify the search path taken during TS depending on the outcome. Having studied numerically the output of the quantum state one can get with QAOA, and the penalty effect, we switch to less idealized simulations where the subproblems depend on the QAOA output during the TS search. 

\vspace{-8mm}
\subsubsection{QAOA exploration possibilities}

\begin{figure}[!ht]
\begin{center}
\vspace{-0.5cm}
\includegraphics[width=0.97\textwidth, trim= 0 2mm 0 0,clip]{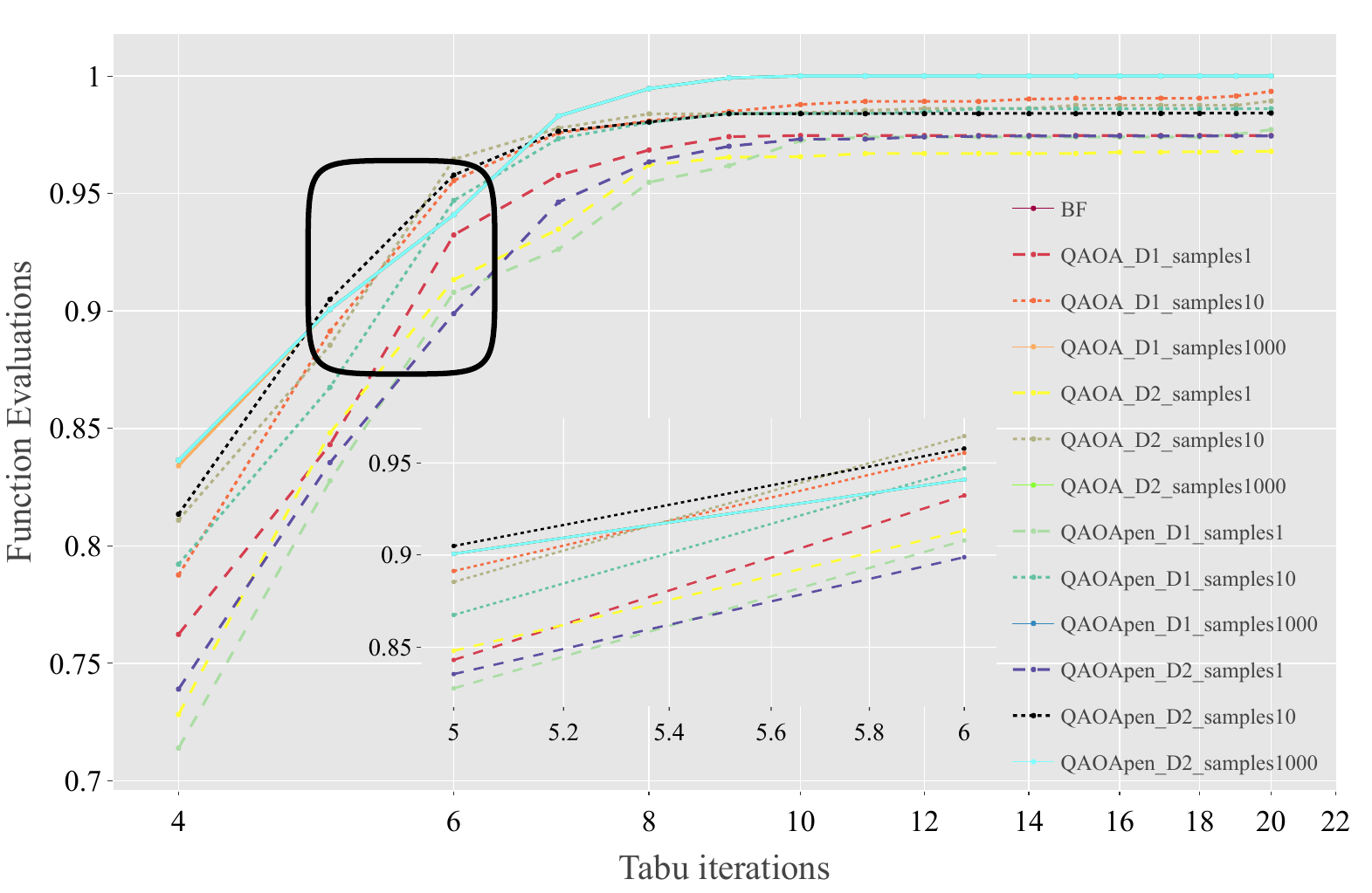}
\caption{Median of best normalized evaluations achieved over TS iterations for instance 1d. The mentions D1, D2, mean respectively $p=1,2$ and the penalized version is indicated by \guillemotleft pen\guillemotright. A higher curve corresponds to better solutions reached. At iteration $20$, the basic TS value would be $0.38$, while the lowest QAOA curve value is $0.967$. The $m=1000$ runs and BF are over $0.99$ starting at the 8th iteration. At iteration $5$, the penalized $p=2, m=10$ version is slightly better than the others, even BF ($0.9049$ vs $0.9007$). At the 6th, the $m=10$ versions are above BF (respectively $0.9645$ and $0.9578$ for unpenalized and penalized $p=2$, $0.9555$ and $0.9470$ for $p=1$, and $0.9409$ for BF).}
\label{sampling1d}
\vspace{-3mm}
\end{center}
\end{figure} 

\begin{figure}[!ht]
\begin{center}
\vspace{-0.5cm}
\includegraphics[width=0.96\textwidth, trim= 0 2mm 0 0,clip]{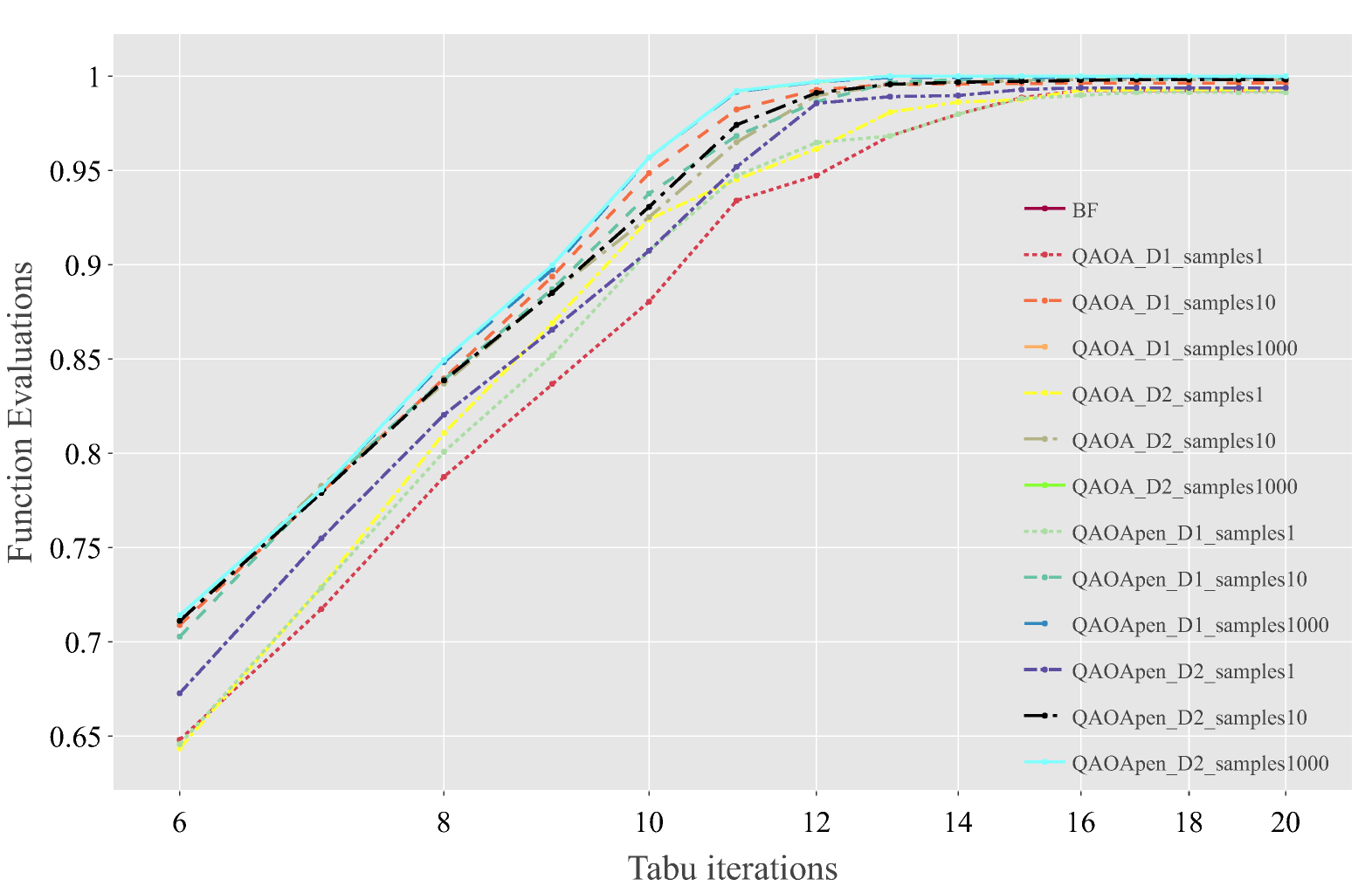}
\caption{Median of best normalized evaluations achieved over TS iteration for instance 1e. At iteration $20$, the lowest QAOA curve value is $0.991$ (while $0.19$ for the basic TS). The $m=1000$ runs and BF are over $0.99$ starting at the 11th iteration. At iteration $7$, the original $p=2$ QAOA using $10$ samples point is higher than the others, even BF ($0.7827$ vs $0.7808$). This also happens at iteration $12$, with the penalized $p=2, m=1000$ ($0.997085$ vs $0.996902$).}
\label{sampling1e}
\end{center}
\end{figure}

After looking at examples of QAOA output and outlining a few possible exploration opportunities, we carried out a few extra simulations but not considering the subproblems obtained with brute-force. Hence, QAOA (and its penalized version) was called once per iteration with BIPOP-CMAES \cite{pycma,bipopcma} optimizing from one set of angles \footnote{When using BIPOP-CMAES, we run circuits with 1000 measurements to estimate expectation values. The optimizer stops when it has reached 2000 evaluations. We obtained great performances in terms of averaged ratios (as the evaluations divided by the optimum of the subproblem), superior to 0.97 at the considered depths.}. In the following, we give a few examples to illustrate the exploration possibilities.
\par 
We take instances where BF simulations required few iterations and where the basic TS was beaten in target. Namely, instance 1d for $k=15$ and $TT=5$ and 1e for $k=15$ and $TT=10$. We run Alg.~\ref{alg:tsoneflipQAOA} for $10$ times, limiting them to 20 TS iterations. Different numbers of samples are used for generating a new candidate per iteration: just once, 10 times and 1000 times. The latter could be considered as a quasi-brute-force approach in these runs. We denote as $m$ the number of samples used in the following.
\par 
Fig.~\ref{sampling1e} and Fig.~\ref{sampling1d} show the median of the best normalized evaluation obtained per run by iteration. We observe from iteration 7 for 1d and 10 for 1e that $m=1000$ is equivalent (in median) to the BF generation. In general, the more samples used, the better the solution found. However, we had at a few iterations median runs that achieved higher values than BF. For instance, this happened for the penalized $p=2$ QAOA at the 5th iteration with $m=10$ for 1d and $m=1000$ at the 12th for 1e. We consider also the frequency of runs for which the basic TS was beaten, and the optimum was found. Table~\ref{res-tablefreqQAOA} summarizes our results. Increasing $m$ improves the frequency of successful runs. We observe that new paths were found, mainly with an extra iteration or two but exceptionally one run or two over 10 could save one iteration. This was the case on instance 1e with 12 iterations instead of $13$, exclusively with the penalized version. These examples are numerically in favor of a greedy (or quasi) approach in solving subproblems. However, QAOA allows, through a trade-off between exploration and exploitation, discovering new paths towards optimality that are still interesting in terms of number of iterations. 

\begin{table}[!ht]
\caption{\small
Frequency of successful runs in beating the basic TS and finding the optima for instance 1d for $k=15$ and $TT=5$, and instance 1e for $k=15$ and $TT=10$ (separated by /), in terms of QAOA settings (with the number of measurements noted $m$). We report also the number of iterations that led to the optimum. }
\vspace{-0.7cm}
\label{res-tablefreqQAOA}
\begin{center}
\begin{scriptsize}
\begin{sc}

\begin{tabular}{l|c|c|c|c}
\toprule
QAOA & $m$ & \multicolumn{1}{|p{3cm}|}{\centering Frequency\\ beating TS (/$10$)} & \multicolumn{1}{|p{3cm}|}{\centering Frequency\\ optimum (/$10$)} & \multicolumn{1}{|p{3cm}|}{\centering Iterations \\to optimum} \\
\midrule

D1 & \multirow{4}{*}{$1$} & 0 / 0 & 0 / 0 &  \\
\cmidrule{3-5}
D1pen& & 0 / 0 & 0 / 0 &  \\
\cmidrule{3-5}
D2& & 0 / 1 & 0 / 1 & / 14\\
\cmidrule{3-5}
D2pen && 0 / 0 & 0 / 0 &  \\
\midrule

D1 & \multirow{4}{*}{$10$} & 7 / 2 & 1 / 2 & 18 / 15,16 \\
\cmidrule{3-5}
D1pen & & 4 / 2 & 0 / 2 & / 15,15\\
\cmidrule{3-5}
D2 & & 5 / 1 & 2 / 1 & 11 / 14 \\
\cmidrule{3-5}
D2pen & & 4 / 0 & 1 / 0 & 10 / \\
\midrule

D1 & \multirow{4}{*}{$1000$} & 8 / 7 & 8 / 7 & 9,10,12 / 13 \\
\cmidrule{3-5}
D1pen &  & 8 / 5 & 8 / 5 & 10 / 12,13 \\
\cmidrule{3-5}
D2 & & 9 / 9 & 9 / 9 & 10,11 / 13,14\\
\cmidrule{3-5}
D2pen & & 10/ 7 & 10 / 7 & 9,10 / 12,13\\
\bottomrule
\end{tabular}
\end{sc}
\end{scriptsize}
\end{center}
\end{table}

\section{Conclusion and outlook}
\label{discussion}

In this work, we studied sampling aspects when quantum approaches, specifically the QAOA algorithm, are considered in combinatorial optimization. We considered a practically relevant setting where a gate-based quantum algorithm, limited in the number of qubits, is utilized in a hybrid quantum-classical framework to solve large optimization instances faster. Our framework constitutes a powerful yet simple heuristic, Tabu Search, in tandem with QAOA as a local neighborhood sampler. 
\par
As a starting point, numerical experiments over open-source QUBO problems up to 500 variables validate using QAOA as a proxy to explore larger neighborhood, under the assumption that subproblems are solved optimally. Continuing, we investigated the exploration possibilities given by QAOA output at small depth. User-defined parameters such as depth and number of measurements used to generate a candidate, can be increased to favor exploitation. On our examples, solving subproblems emphasizing more on the latter gave better general performances. Yet, we found that exploration can be beneficial. Iterations can be saved with our QAOA procedure, illustrating that missing to generate the solution of a subproblem in previous iterations could yield to faster paths towards better solutions. Hence, the QAOA-based algorithm we introduce in this work becomes a very flexible tool in such hybrid quantum-classical settings. 
\par
We see numerous possibilities for future work. First, our model allows for many hyperparameters whose function needs to be explored, and, as is usually done in many local search methods, the exploration/exploitation trade-offs can be made online-adaptive. Second, the effect of real world limitations, most importantly noise, and hardware connectivity, calls for further investigation. Although QAOA can be run on real hardware \cite{qaoasycamore,qaoaibm},  its output quality will improve as the quantum devices decreases in error, or through quantum error mitigation \cite{qem}. Finally, it would be interesting to propose different frameworks (e.g. \cite{MTS,ITS,GloverDiversificationdrivenTS,L2010AHM}) with special emphasis on the exploration possibilities given by small-depth quantum algorithms, and cross-compare with standard techniques in future works. We believe our approach combined with these types of analyses will provide new promising ways to maximize the use of limited near-term quantum computing architectures for real world and industrial optimization problems.

\section*{Acknowledgements}

CM, TB and VD acknowledge support from Total. This work was supported by the Dutch Research Council (NWO/OCW), as part of the Quantum Software Consortium programme (project number 024.003.037). This research is also supported by the project NEASQC funded from the European Union’s Horizon 2020 research and innovation programme (grant agreement No 951821).

\bibliography{references}
\bibliographystyle{splncs04}

\appendix
\section*{Appendix}

\begin{figure}[p]
\begin{center}
\includegraphics[width=0.8\textwidth]{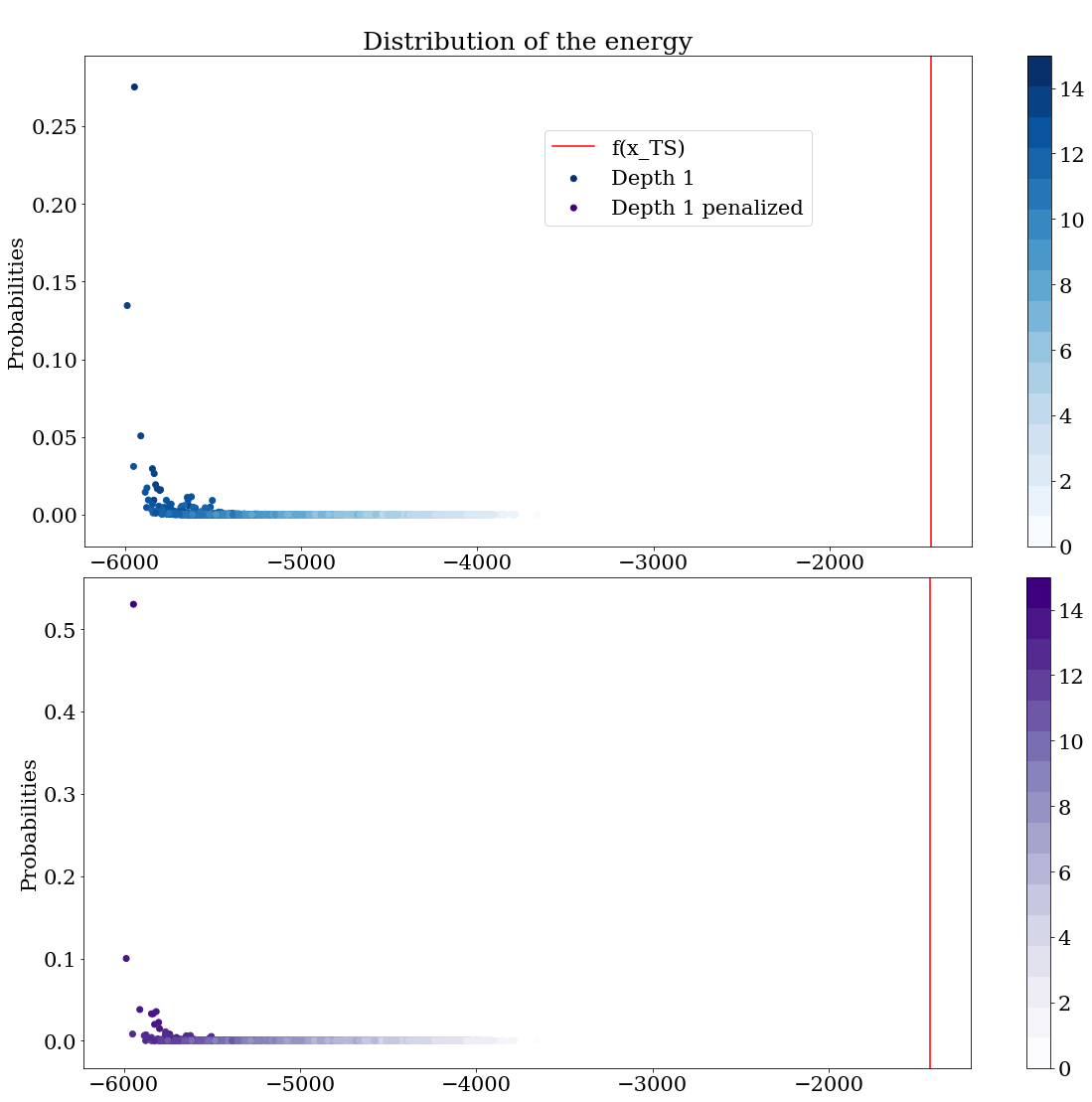}
\includegraphics[width=0.8\textwidth]{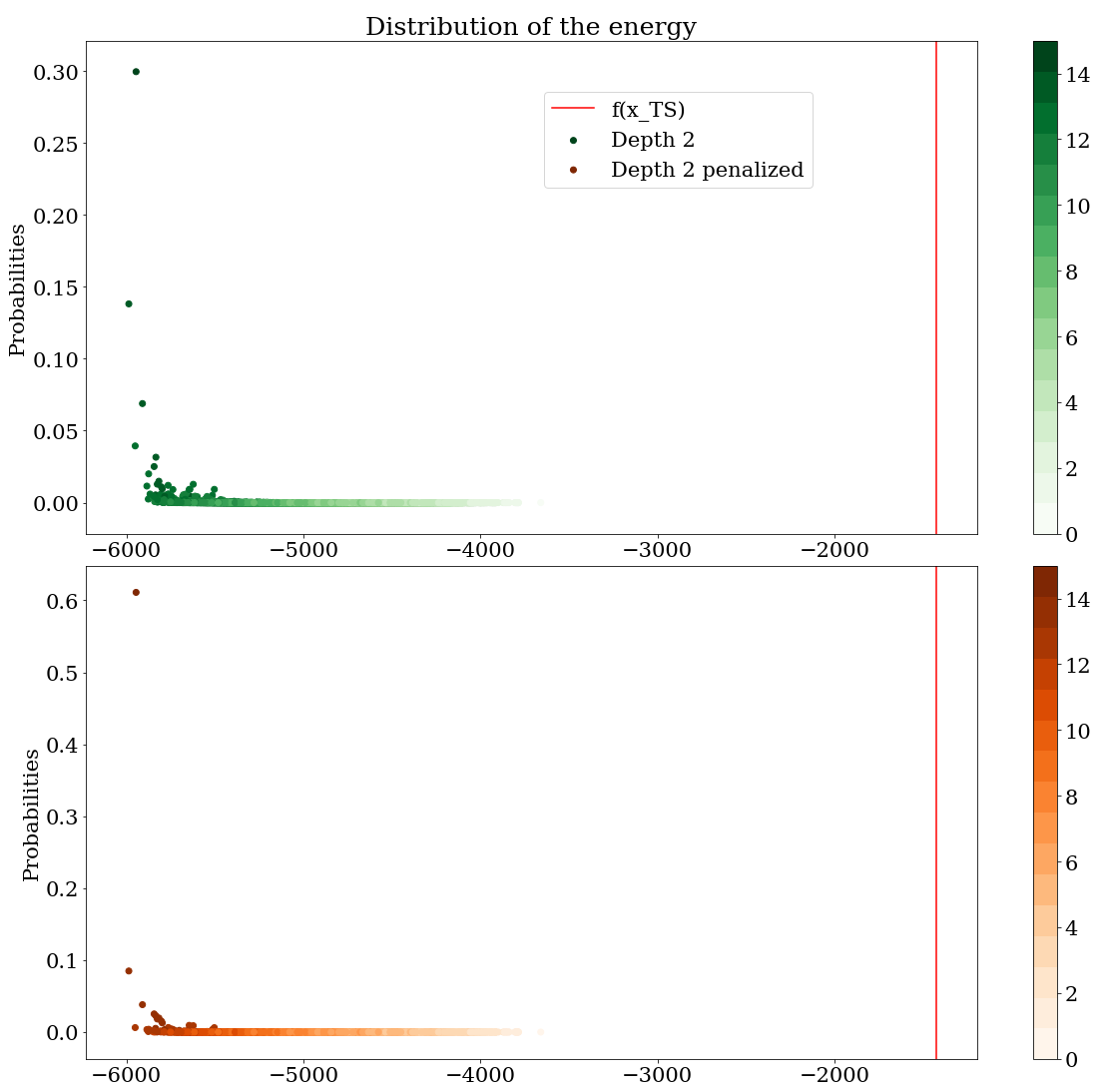}
\caption{Distribution of the evaluations (or energies) for the third iteration obtained on instance 1e, given by the QAOA output state at depth $1$ and $2$, with and without the penalty term. A colormap is given for the Hamming distance with the current tabu solution.}
\label{qaoa_penalized_dist_greedy}
\end{center}
\end{figure} 

\begin{table}[!p]
\caption{Best evaluations obtained for $TT=5,10,15$ with the first TS iteration to reach the corresponding maximum given in~\cite{amts}. Basic TS is run for 20000 iterations. The other algorithms are run for 1000 iterations. The brute-force approach with $k=20$ is run for 200 iterations. The best performances per instance and tabu tenure are highlighted in bold.}
\label{res-tabledef}
\vskip -0.15in
\begin{center}
\begin{scriptsize}
\begin{sc}
\begin{tabular}{l|c|l|l|l|l|l}
\toprule
Algo. & TT & 1d (6333) & 2d (6579) & 3d (9261) & 4d (10727) & 5d (11626) \\
\midrule

Basic & \multirow{4}{*}{5}  & 6272 / 64 & 6579 / 71 &  9261 / 90 & 10727 / 103 & 11613 / 109 \\
\cmidrule{3-7}
$k=10$  &&  6328 / 13 & 6579 / 160 & 9261 / 73 & 10727 / 43 & \textbf{11626 / 19} \\
\cmidrule{3-7}
$k=15$  &&  6333 / 10 & \textbf{6579 / 26} &  9261 / 8 & 10727 / 22 & 11626 / 42 \\
\cmidrule{3-7}
$k=20$  &&  \textbf{6333 / 7} & 6579 / 37 &  \textbf{9261 / 5} & \textbf{10727 / 11} & 11613 / 9 \\

\midrule
Basic  & \multirow{4}{*}{10}  & 6333 / 658 & \textbf{6579 / 71} &  9261 / 485 & 10714 / 91 & 11626 / 149\\
\cmidrule{3-7}
$k=10$  &&  6333 / 568 & 6579 / 121 & 9247 / 294 & 10727 / 72 & \textbf{11626 / 19} \\
\cmidrule{3-7}
$k=15$  &&  6333 / 10 & 6579 / 287 & 9261 / 8 & 10714 / 685 & 11626 / 988 \\

\cmidrule{3-7}
$k=20$  &&  \textbf{6333 / 7} & 6531 / 149 &  \textbf{9261 / 5} & \textbf{10727 / 11} & 11613 / 9 \\

\midrule
Basic  & \multirow{4}{*}{15}  & 6333 / 300 & \textbf{6579 / 71} & 9261 / 750 & 10727 / 545 & 11626 / 891 \\
\cmidrule{3-7}
$k=10$  &&  6328 / 13 & 6579 / 560 &  9261 / 422 & 10684 / 822 & \textbf{11626 / 19} \\
\cmidrule{3-7}
$k=15$  &&  6333/ 10 & 6533 / 12 &  9261 / 8 & 10686 / 545 & 11589 / 7 \\
\cmidrule{3-7}
$k=20$  &&  \textbf{6333 / 7} & 6484 / 141 &  \textbf{9261 / 5} & \textbf{10727 / 11} & 11613 / 9 \\
\bottomrule

\addlinespace[0.2cm]

\toprule
Algo. & TT & 1e (16464) & 2e (23395) & 3e (25243) & 4e (35594) & 5e (35154) \\
\midrule
Basic    & \multirow{4}{*}{5}  & 16410 / 132 & 23329 / 199 &  25228 / 289 & 35578 / 147 & 35126 / 219 \\
\cmidrule{3-7}
$k=10$  && 16439 / 19 & 23330 / 138 & 25243 / 21 & 35594 / 111 & \textbf{35154 / 25} \\
\cmidrule{3-7}
$k=15$  &&  \textbf{16464 / 13} & \textbf{23395 / 115} &  \textbf{25243 / 13} & 35578 / 66 & 35154 / 58\\

\cmidrule{3-7}
$k=20$  &&  16458 / 85 & 23330 / 12 &  25236 / 54 & \textbf{35594 / 69} & 35154 / 135 \\

\midrule
Basic  & \multirow{4}{*}{10}  & 16458 / 215 & 23323 / 144 &  25243 / 190 & 35578 / 147 & 35149 / 1068 \\
\cmidrule{3-7}
$k=10$  &&  16458 / 323 & 23395 / 254 & 25243 / 24 & \textbf{35594 / 122} & \textbf{35154 / 25} \\
\cmidrule{3-7}
$k=15$  &&  \textbf{16464 / 13} & \textbf{23395 / 34} &  \textbf{25243 / 13} & 35579 / 304 & 35154 / 226\\

\cmidrule{3-7}
$k=20$  &&  16425 / 176 & 23372 / 104 & 25243 / 146 & 35578 / 12 & 35102 / 100\\

\midrule
Basic    & \multirow{4}{*}{15}  & 16464 / 4107 & \textbf{23395 / 4253} &  25243 / 500& \textbf{35594 / 170} & 35126 / 2683 \\
\cmidrule{3-7}
$k=10$  && 16439 / 19 & 23381 / 897 & 25243 / 21 & 35538 / 348 & \textbf{35154 / 25} \\
\cmidrule{3-7}
$k=15$  &&  \textbf{16464 / 13} & 23389 / 788 &  \textbf{25243 / 13} & 35559 / 795 & 35126 / 305 \\
\cmidrule{3-7}
$k=20$  && 16415 / 187 & 23330 / 12 &  25124 / 140 & 35578 / 12 & 34986 / 33\\
\bottomrule

\addlinespace[0.2cm]
\toprule
Algo. & TT & 1f (61194) & 2f (100161) & 3f (138035) & 4f (172771) & 5f (190507) \\
\midrule
Basic  & \multirow{4}{*}{5}  & 60770 / 406 & 99959 / 498 &  138031 / 464 & 172449 / 510 & 190498 / 510 \\
\cmidrule{3-7}
$k=10$  &&  60980 / 92 & 100047 / 228 & 138035 / 183 & 172449 / 127 & 190502 / 126 \\
\cmidrule{3-7}
$k=15$  &&  60988 / 43 & \textbf{100161 / 272} &  138021 / 697 & 172734 / 46 & 190406 / 90 \\
\cmidrule{3-7}
$k=20$  &&  \textbf{61057 / 179} & 100158 / 101 &  \textbf{138035 / 61} & \textbf{172748 / 186} & \textbf{190502 / 74} \\

\midrule
Basic   & \multirow{4}{*}{10}  & 61189 / 746 & 99850 / 1317 &  138035 / 631 & 172391 / 834 & 190498 / 720 \\
\cmidrule{3-7}
$k=10$  &&  61087 / 310 & \textbf{100150 / 564} &  137988 / 83 & 172346 / 697 & 190445 / 587 \\
\cmidrule{3-7}
$k=15$  &&  \textbf{61194 / 444} & 100136 / 543 &  \textbf{138035 / 139} & \textbf{172734 / 46} & \textbf{190502 / 593} \\

\cmidrule{3-7}
$k=20$  &&  61087 / 184 & 99788 / 177 & 137967 / 112 & 171991 / 120 & 190373 / 117 \\

\midrule
Basic   & \multirow{4}{*}{15}  & 61193 / 1450 & \textbf{100161 / 5337} & 138035 / 927 & 172734 / 1148 & \textbf{190502 / 2045} \\

\cmidrule{3-7}
$k=10$ &&  \textbf{61194 / 213} & 100120 / 579 & \textbf{138035 / 77} & 172413 / 86& 190320 / 796 \\

\cmidrule{3-7}
$k=15$  &&  61178 / 998 & 100043 / 477 & 137890 / 912 & \textbf{172734 / 46} & 190374 / 517 \\
\cmidrule{3-7}
$k=20$ &&  60947 / 90 & 100051 / 179 & 137657 / 134 & 172309 / 190 & 189588 / 26 \\

\bottomrule
\end{tabular}
\end{sc}
\end{scriptsize}
\end{center}
\vskip -0.1in
\end{table}

\end{document}